# Development of a 3D virtual world tool for sustainable energy education

**Marta Guerra-Mota[a], Dimosthenis Minas[b], Michalis Xenos[b] and Maria Manuel Sá[c]**

[a]Department of Business Sciences and UNICES, University of Maia, Portugal

[b]Department of Computer Engineering and Informatics, University of Patras, Greece

[c]Department of Business Sciences, University of Maia, Portugal and NECE-UBI, Research Centre for Business Sciences, 6200-209 Covilhã, Portugal

## Abstract

The UNESCO (2022) points out that the gap between the existing awareness of a person or a community, and the actual habits of everyday life, is attributed to: low levels of understanding the environmental issues at stake; low levels of knowledge regarding energy and climate issues; and lack of attention to social, emotional or behavioral learning. In this context, as environmental and energetic concerns grow rapidly in our daily lives, RAISE - Raising environmental knowledge & awareness, an ERAMUS+ project with partners from three countries (Greece, Portugal, and Italy), was born. The project's first and main objective is to increase the environmental knowledge and awareness (EK&A) of schoolchildren. To achieve this goal, a survey was first carried out to ascertain the students' knowledge and, based on these results, scenarios and educational material were drawn up to build a pedagogical tool - a 3D Virtual World Learning Environment (3D VWLE). This paper presents the desktop research results and the main features of the game that are intended to match the identified educational needs. The 3D VWLE puts students in situations where they can acquire transferable skills. The project demonstrated that the simulation effect of 3D VWLE on skills provides an excellent online learning environment for students to work on and improve their abilities. In other words, virtual worlds open the door to a new way of learning and teaching.

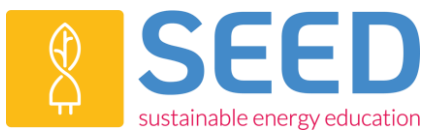

## 1. Introduction

It is widely acknowledged that energy consumption has become a critical facet of modern living; a significant proportion of individuals and organizations now dedicate considerable resources to managing energy use, exploring renewable sources, and leveraging technology for more efficient energy consumption. Despite the benefits that come with advancements in energy technologies, such as reduced greenhouse gas emissions, increased energy security, and economic savings, there are also notable challenges and negative impacts associated with our current energy practices, including environmental degradation, resource depletion, and the exacerbation of climate change. The complexity of the energy landscape necessitates informed decision-making; education emerges as a pivotal tool for enhancing understanding and awareness around critical energy issues. In response to this need, this paper introduces an innovative educational approach designed to enrich knowledge and implementation of this educational initiative, the paper aims to contribute significantly to improving awareness and actions regarding sustainable energy practices, highlighting the essential role of education in navigating the energy challenges of the 21$^{st}$ century.

This educational tool represents an advanced online, three-dimensional virtual world designed to encompass a broad spectrum of energy education concepts, including but not limited to wind farm management. It serves as an immersive platform where users can engage with various energy-related scenarios, ranging from renewable energy sources like wind and solar, to the intricacies of energy efficiency, grid management, and the environmental and societal impacts of energy choices. Through interactive simulations, users encounter real-life challenges such as optimizing renewable energy production, balancing energy demands with supply, and mitigating the negative effects of energy consumption on the environment.

The primary goals of this paper are to introduce this comprehensive educational tool, detail the diverse scenarios that have been developed to cover a wide range of energy topics, discuss the methodologies employed for evaluating its effectiveness, and share the outcomes achieved from its implementation. Additionally, the paper explores how this three-dimensional virtual world contributes to the learning process, enhancing the educational journey for students enrolled in pre-university studies.

Research in learning indicates that hands-on learning is an active and constructive process and becomes more impactful when rooted in real-world learning styles (Bransford, 2000); methodologies such as problem-based learning, cooperative learning, and activity-led learning can significantly enhance students' critical thinking skills. Moreover, effective learning arises from the interaction between an actively engaged learner and a supportive, stimulating learning environment (Lombardo, 2009). In this context, the 3DVW (Three-Dimensional Virtual World) introduced in this paper provides the ideal infrastructure to offer learners a comprehensive understanding of various situations through immersive experiences, facilitating a hands-on approach to learning that leverages the engaging and interactive potential of virtual environments.

The development and use of 3D Virtual World Learning Environment (3D VWLE) for educational purposes represents a dynamic area that expands and enriches the concept of educational spaces. In such environments, learners have the freedom to navigate, explore, engage in activities, make errors, and collaborate or communicate with peers. The primary goal within a 3D VWLE is to foster a sense of immersion, allowing learners to feel as though they are truly part of the environment. This immersion, combined with interaction with virtual elements, can significantly boost learners' interest and engagement in educational tasks, potentially leading to a deeper understanding of concepts contingent on the subject manner.

Within this immersive 3DVWLE, users are transported into an engaging virtual setting where they encounter real-life energy threat scenarios. These scenarios are ingeniously presented as interactive games, allowing users to navigate through, engage with, and learn about various aspects, risks and threats associated with energy use. The design of this educational tool leverages the immersive and interactive capabilities of 3DVWLEs to simulate experiences that mirror real-world challenges. However, it ensures that learners can explore these serious issues in a manner that is both safe and enjoyable.

The educational tool's focus on energy saving and addressing energy threats through a 3D VWLE represents a significant step forward in environmental education. It showcases how advanced digital technologies can be harnessed to tackle critical issues, making complex concepts more accessible and engaging for a broader audience (Grivokostopoulou et al. 2020). This approach not only educates but also empowers individuals with the knowledge and experience needed to make informed decisions and act towards a more sustainable future (An, 2019; Lo, 2022).

## 2. Methodology

The methodology was divided into three main phases. Firstly, desktop research was performed, to identify critical issues in the energy and environment fields, based on a literature review and analysis of reports from reference organizations. The creation of a questionnaire to assess the knowledge, behavior, and awareness of school pupils relating energy and environmental issues. This activity was sourced on the results of the previous phase and on a review of the literature on education for sustainability. The data obtained of the respondents was subjected to statistical analysis and discussion to support the scenarios proposals. Finally, the scenarios were developed, which, through a narrative in which the player is involved, enable the proposed learning objectives to be met.

The questions from the survey were elaborated after consulting international scientific articles on the topic (Michalos et al., 2012; Gericke et al., 2018; Lestari et al., 2022). Although these authors only used questions with a Likert scale response, the project team chose to also pose multiple choice questions and questions with optional answers. Thus, the block consists of 21 questions with optional answers and one multiple choice question; the Attitudes block consists of 4 questions with optional answers and 6 questions on a Likert scale (from 1 Strongly disagree to 5 Strongly agree); the Behavior block consists of 24 questions with Likert scale answers (from 1 Never to 5 Often). The survey consisted of a sociodemographic set of questions (with the variables age, gender, country, and school year) and the three blocks with questions about knowledge, attitudes and behaviors regarding energy and climate issues. The questionnaire was carried out using the software LimeSurvey in English, Greek, Portuguese, and Italian versions. The data has been collected through an online survey in three basic and secondary schools of the partners involved in the research. Students from Arseiko Lyceum of Patra from Greece, Agrupamento de Escolas D. Afonso Sanches from Portugal, and Liceo Manin from Italy. The criteria defined for the inclusion of participants in the sample were: (a) students from basic and secondary school; (b) female and male gender; (c) acceptance to participate voluntarily in the study with the informed consent of the parents. The data collection instruments were administered to the students online via the schools' computers. The questionnaires were previously validated by a group of 10 students in each school to identify opportunities for improvement or difficulties in interpreting the questions. As no issues to be changed were identified, the questionnaires were then disseminated to a larger group of students. The sampling and data collection process took place between the 1st and 31st of March of 2023. Data was processed using version 28.0 of the Statistical Program for Social Sciences (IBM SPSS). The results obtained were subjected to a statistical analysis which identified the critical areas in terms of knowledge and attitudes towards environmental and energy issues, which supported the scenario development of the virtual world.

For the development of the 3D Virtual World (3DVW), a comprehensive survey of available open-source virtual world platforms was undertaken (Maratou et al., 2015). OpenSimulator was selected for its robust features, including a built-in 3D Editor, scripting capabilities, and server infrastructure. This choice enabled the crafting of dynamic, fully interactive multi-user 3D environments from the ground up. OpenSimulator's architecture facilitates the creation of immersive virtual spaces where users can engage in a variety of educational activities, enhancing the learning experience with its interactive and engaging virtual environments.

## 3. Results and discussion

Energy issues cut across a wide range of topics, from technology, greenhouse gas emissions, consumption and production, transport, among others, which were addressed in the survey. It was intended to be comprehensive to cover different energy-related topics that could support interesting and challenging scenarios.

Regarding the questions in the Knowledge Section (Save energy): the renewable energy most known to students is solar energy (95.1%) followed by wind energy (92.2%). The least known renewable energy is biomass (52%) followed by ocean energy (54.6%). This shows that there are renewable energy alternatives that are little known to young people and/or that some forms arouse greater interest, such as wind or solar energy. Regarding the Behavior section, in general, students recycle and engage in energy-saving behavior. However, 15 to 20% do not encourage their family and/or friends to do the same. Also, more than 50% of the students never or rarely get involved in activities related to gardening and growing fruits and vegetables. Although students consider it a positive attitude to join an environmental group, 58.5% of students never collaborate with an environmental group that works to protect the environment and promote sustainability. Only 37,7% of participants always or often share the car with friends or neighbors. 61,1% of the participants have never taken part in a protest action in defense of the environment. Given that today's youth are extremely connected to social networks 79,2% have never or rarely published posts on social media about environmental causes.

In general, young respondents perform well in the survey but also present opportunities to improve their behavior, attitudes and knowledge regarding the topics covered. Although attitudes score excellent in most cases, the actions do not always accompany them, so tools to support better performances and greater interventions are desirable. Building upon the solid foundation provided by OpenSimulator, the development team designed a suite of educational simulations aimed at embedding key sustainability and environmental concepts within interactive 3D virtual worlds (3DVW). These simulations were engineered to leverage experiential learning, employing immersive, scenario-based activities that place users in various situations where their decisions have direct implications on environmental outcomes. Through engaging in these virtual experiences, users are prompted to apply critical thinking and problem-solving skills, mirroring the complexities of real-world environmental management and personal responsibility towards sustainability.

The design approach for these simulations utilized branching scenarios, a storytelling method that offers users multiple pathways based on their decisions, to simulate the consequences of their actions in a virtual setting (Siordia-Medina, 2020). This methodology not only makes the learning process more engaging but also personalizes the educational experience (Thongsri et al.2019), allowing users to see the direct impact of their choices. Such an approach was instrumental in conveying the intricate balance between daily human activities and their environmental footprints, as well as the strategic considerations involved in managing renewable energy resources. By integrating these themes within captivating and interactive games, the simulations aimed to foster a deeper understanding of sustainability practices and encourage a reflective attitude towards personal and collective environmental impact.

*A. Wind Farm Challenge*

The "Wind Farm Challenge" scenario in the game (figure 1) is designed as an educational and interactive single-player simulation. This scenario aims to teach users about the aspects and challenges of wind energy production and wind farm management. The player takes on the role of a participant in the Wind Farm Challenge, where they are tasked with designing an efficient and sustainable wind farm within specific environmental and budgetary constraints. The game includes interactive elements such as discussions with Non-Player Characters (NPCs or 'bots') experts, a quiz on wind energy, and tasks that require critical thinking and decision-making based on environmental impact and technical considerations. The educational content is woven into the gameplay, providing an engaging learning experience about renewable energy and sustainability.

**Fig. 1 – Educational scenario about wind energy - Wind farm challenge**

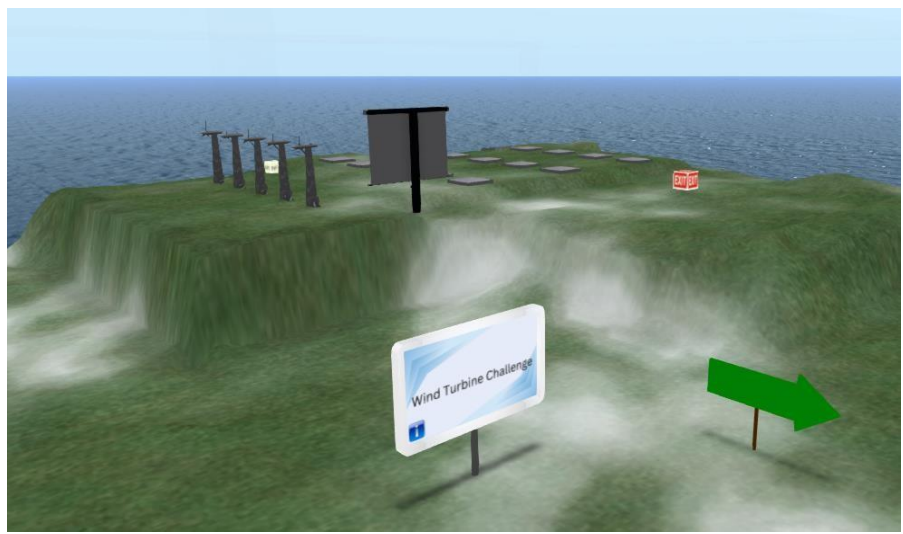

Source: Elaborated by the authors (2024)

*B. Carbon Champions*

The "Carbon Champions" scenario is an educational game designed to teach players about carbon footprints and sustainable living (figure 2). Set in everyday environments like a home or school, players take on the role of a student making daily choices that impact their carbon footprint. The game features a carbon calculator that reflects the environmental impact of these choices. Players engage in activities like selecting meals, modes of transport, and energy use, each affecting their carbon footprint. The aim is to maintain a low carbon footprint, with the game providing feedback and information on how each choice contributes to or mitigates climate change. This interactive and informative scenario encourages players to reflect on and adjust their habits for a more sustainable lifestyle.

Fig. 2 – Educational scenario about carbon footprint – Carbon champions

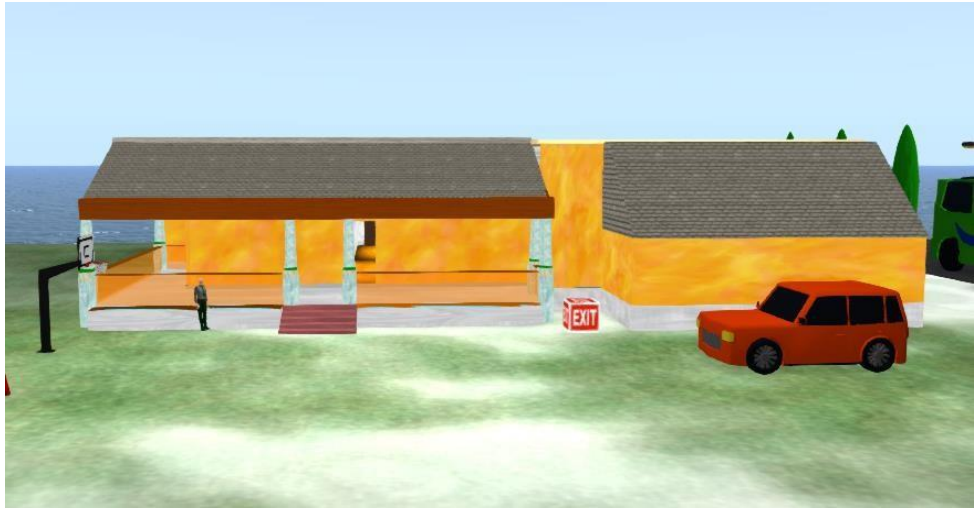

Source: Elaborated by the authors (2024)

*C. Other Areas of the Environment*

The 3D Virtual World (3DVW) comprises diverse virtual environments, each hosting a unique game (example in figure 3). These environments are populated with interactive 3D elements like doors and chairs, enabling avatar interaction. They also feature NPCs, which are computer-generated avatars participating in or enriching the scenarios. Additional features include text communication systems, user-friendly navigation, avatar performance monitoring, scoring mechanisms, feedback provision, tailored assistance, and methods for evaluating knowledge.

Fig. 3 - Educational scenario about recycling- Lost in the waste.

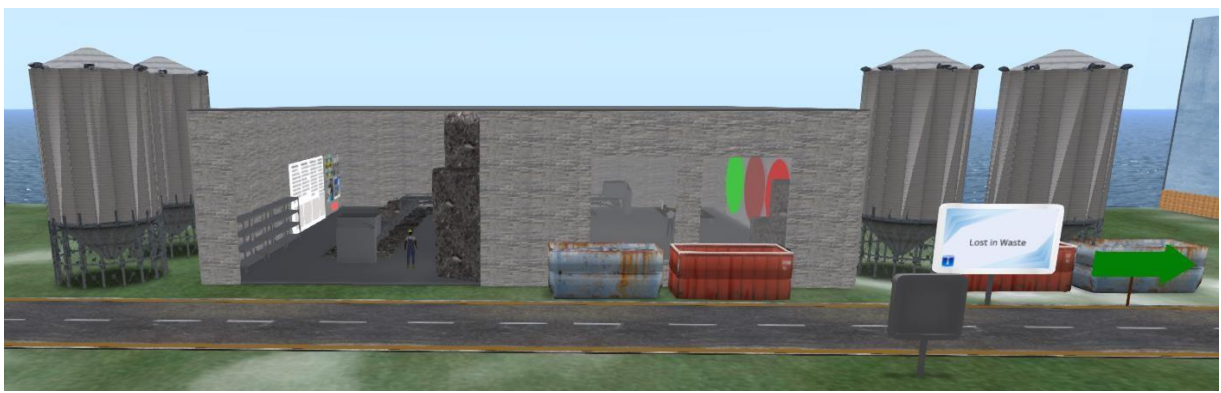

Source: Elaborated by the authors (2024)

Within the 3D Virtual World, the "Welcome Area" serves as a hub for users, featuring a scenario selection room where they can choose from 10 different environmental scenarios. Each scenario is uniquely designed to raise awareness about a specific environmental issue. Additionally, there's a tutorial room where users learn basic game mechanics, such as movement and interaction with NPCs. This setup is structured to facilitate user engagement and education in various environmental topics through interactive and immersive virtual experiences. The 'in-world' textual content and the educational material of the 3DVW is produced and implemented in five languages: English, Greek, Italian and Portuguese.

## 4. Conclusions

To achieve the proposed goals, an innovative tool for teaching and learning practices was developed, changing learning techniques from teacher-centered to student-centered. Active learning helps students to rethink learning as a creative, engaging, and constructive process. The pedagogical tool, a 3Dimension Virtual World Learning Environment (3D VWLE), allows the simulation of activities with gamified elements that educate students through action; knowledge will be perceived not only through text, but through multiple modalities, while students' environmental awareness and knowledge will be reinforced by a visual perception as close to reality as possible. For the next steps, we expect to collect additional information from the players, in school and on laboratory environment, which will allow us to assess the effective impact of the tool on learning.

## Acknowledgements

This project is co-funded by Erasmus+ Programme of European Union, action KA220-SCH Cooperation partnerships in school education, under the contract 2022-1-EL01-KA220-SCH-000088295. The European Commission's support for the production of this publication does not constitute an endorsement of the contents, which reflect the views only of the authors, and the Commission cannot be held responsible for any use which may be made of the information contained therein.

## Conflicts of interest

The authors declare that they have no known competing financial interests or personal relationships that could have influenced the work reported in this paper.